\documentclass[a4paper]{amsart}
\usepackage{amssymb,amsthm,euscript,amsmath,graphicx,array}

\newtheorem{thm}{Theorem}[section]
\newtheorem{lem}[thm]{Lemma}

\newtheorem{rem}{Remark}[section]

\def\rev{^{{\rm r}}}  % reversible
\def\drho{\partial_\rho} % one variable derivation
\def\visc{^\text{v}}    %viscous
\def\fl#1{_\text{#1}}
\def\Euler{\fl{Euler}}  %Euler
\def\SchM{\fl{SchM}}    %Schrödinger-Madelung
\begin{document}

\pagestyle{plain}

%---------------------------------------------------------------
\title{Weakly nonlocal non-equilibrium thermodynamics -
   variational principles and {S}econd {L}aw}
\author{P\'eter V\'an}
\address{Department of Theoretical Physics, KFKI, Research Institute of Particle and Nuclear Physics,\\ 
Konkoly Thege Mikl\'os \'ut 29-33., 1525 Budapest, Hungary \\
Department of Energy Engineering, Budapest University of Technology and Economics\\
Bertalan Lajos u. 4-6. 1111 Budapest, Hungary}

\email{vpet@rmki.kfki.hu}

\date{\today}

\begin{abstract}
A general, uniform, rigorous and constructive thermodynamic approach to weakly nonlocal non-equilibrium thermodynamics is reviewed. A method is given to construct and restrict the evolution equations of physical theories according to the Second Law of thermodynamics and considering weakly nonlocal constitutive state spaces. The evolution equations of internal variables, the classical irreversible thermodynamics and Korteweg fluids are treated. 
\end{abstract}

\dedicatory{Dedicated to the 70th birthday of J\"uri Engelbrecht.}
\maketitle
%---------------------------------------------------------------

\section{Introduction}

Weakly nonlocal, coarse grained, phase field and gradient are
attributes of theories from different fields of physics indicating
that in contradistinction to the traditional treatments, the
governing equations of the theory depend on higher order space derivatives of the state variables. The origin of idea goes back to the square gradient model of van der Waals for phase interfaces \cite{Waa894a}, where it is extensively applied \cite{BedAta03a,JohBed03a,JohBed04a,KjeBed08b}. Later applications go far beyond phase boundaries or thermodynamics. Nowadays weakly nonlocal is a nomination in
continuum physics dealing with internal structures \cite{Grm93a,LebGrm96a,Mar02a,Mau06a,CimFri07a,Mor08a},
coarse grained or phase field appears in statistically motivated
thermodynamics \cite{HohHal77a,PenFif90a,BroSpr96b,AraKra02a,BedAta03a}, and gradient is
frequently used in mechanics in different context \cite{Mau79a,Mau80a,Mau90a1,VarAif94a,KosWoj95a,Kos96p,Val96a,CimKos97a,Val98a,Bed00a,IreNgu04a,PapFor06a}. The simplest way to demonstrate the meaning of weakly nonlocal extensions can be exemplified
by the Ginzburg-Landau equation which is not only a specific equation in superconductivity, as it was introduced originally by Landau and Khalatnikov \cite{LanKha54a}, but  a first
weakly nonlocal extension of a homogeneous relaxation equation of
an internal variable. The traditional derivation of the
Ginzburg-Landau equation is based on a characteristic mixing of
variational and thermodynamic considerations. One applies a
variational principle for the static part and the functional
derivatives are introduced as thermodynamic forces into a
relaxation type equation. A clear variational derivation to obtain
a first order differential equation is impossible without any
further ado (e.g. without introducing new variables to avoid the
first order time derivative, which is not a symmetric operator)
\cite{VanMus95a}. One can apply these kinds of arguments in
continuum theories in general, preserving the doubled theoretical
framework separating reversible and irreversible parts of the
equations \cite{GrmOtt97a,OttGrm97a,Ott05b}. However, there are also
other attempts to unify the two parts with different additional
hypotheses and to eliminate this inconsistency of the traditional approach \cite{Gur96a,CapMar01a,Mar02a}.

The ultimate aim is to find a unified, general, rigorous  and predictive theoretical framework that makes possible to extend the governing
equations of physics with higher order gradients of the continuum fields, beyond the traditional terms. 
The method should be uniformly applicable from  classical systems in local equilibrium up to relativistic systems beyond local equilibrium; should be general to incorporate most of the mentioned classical examples of weakly nonlocal theories without specific assumptions; should reduce the independent additional assumption to a minimum and should be constructive to give calculational methods for systematic higher order extensions of the constitutive space. 

This paper is a general tutorial to the mathematical framework of such a theory.
In the first section a general methodology of exploiting the Second Law for weakly nonlocal   systems is given. The Second Law is considered as a constrained inequality, where the constraints are the evolution equations of the system and their derivatives, depending on the order of the nonlocality. At the third section evolution equations of internal variables and their different weakly nonlocal extensions are treated. A first order weakly nonlocal theory leads to relaxation type ordinary differential equations, a second order nonlocality leads to Ginzburg-Landau equation and a second order nonlocal theory to dual internal variables  unifies the evolution equations of internal variables derived by mechanical methods  (by variational principles and dissipation potentials) and by thermodynamics (by heuristic application of the Second Law).  At the fourth section we show that classical irreversible thermodynamics can be incorporated naturally in our treatment, a first order weakly nonlocal theory of balance type evolution equations leads to the thermodynamic flux-force relations of classical irreversible thermodynamics with gradients of the intensives as thermodynamic forces. Finally we   demonstrate the applicability of the method to one component heat conducting Korteweg fluids, that  are first order weakly nonlocal in the energy and in the velocity  and second order weakly nonlocal in the density. In that case nontrivial forms of the pressure tensor ensure the compatibility to the Second Law. As a particular example we derive the constitutive functions of the Schr\"odinger-Madelung fluids. Finally  a summary and discussions follow.

\section{Second law and  weakly nonlocal constitutive spaces }

In this section we shortly summarize some methodological specialities of exploiting the Second Law in weakly nonlocal systems. One can find  some further details in \cite{Van01a2,Van03a}, where the role of the entropy flux is treated and in \cite{Van05a}, where the key idea is introduced, the derivative of the constraints as additional constraint. 

There are several interpretations of the Second Law in non-equilibrium thermodynamics (see e.g. \cite{Hut77a,MusEhr96a}). If we wanted to ensure that the Second Law will be a consequence of material properties, then we should postulate that the entropy was an increasing function along the solutions of the evolution equations. With this assumption we are compatible with the classical heuristic method that puts the evolution equations in the balance of the entropy,  and exploits the consequences of the inequality. In the classical approach a quadratic expression is recognized and then a relation is sought between the introduced thermodynamic fluxes and forces \cite{Eck40a3,GroMaz62b}. Here the evolution equations are considered as constraints for the inequality of the entropy production \cite{ColMiz64a}. We arrive to the Coleman-Noll procedure recognizing the algebraic part of the problem where the different derivatives are independent.    When instead of putting constraints into the evolution equations, the constrained algebraic 
inequality  is solved by multipliers then the calculation is called Liu procedure
 \cite{Liu72a}. The Coleman-Noll and Liu procedures are  equivalent in simple systems \cite{TriAta08a}, but the later one preserves the symmetries of the evolution equations and the constraints.  Liu procedure is based on a linear algebraic theorem,
called Liu's theorem in the thermodynamic literature
\cite{Mul72a,MusAta01a} and an interpretation of the role of
entropy inequality.
Hauser and Kirchner recognized that Liu's theorem is a
consequence of a famous statement of optimization theory and
linear programming, the so called Farkas's lemma \cite{HauKir02a}.
That theorem was proved first by Farkas in 1894 and independently
by Minkowski in 1896 \cite{Min896b}. In the Appendix we have formulated and proved the classical Farkas lemma, the affine Farkas lemma and Liu's theorem.

Originally Farkas developed his lemma to formulate correctly the
Fourier's principle of mechanics of mass-points, which is the
generalization of d'Alembert's principle in case of inequality
constraints \cite{Far894a}. The role of Liu's theorem  in
continuum physics is similar in some sense: we want to give
the correct form of the evolution equations taking into account the
requirement of  the Second Law, that is the entropy inequality.
In fact only the material part of the evolution equations - the constitutive functions - is/are restricted. The  mathematical formulation introduces a kind of dual point of view, because in  Liu
procedure the evolution equations - the partial or ordinary differential equations determining the evolution of the system - is a condition, a constraint for the entropy inequality. The variables (e.g. fields) in the evolution equation form the {\em basic state space}. The constitutive quantities
depend on these functions, on the {basic state} and some of its
derivatives. To make the problem algebraically manageable the
basic state variables and their derivatives are considered as
independent quantities. Some of them can be incorporated into the
\textit{constitutive state space} (or large state space
\cite{MusAta01a}), into the domain of the constitutive functions.
The entropy inequality, our objective function, has a special balance form, and determines the {\em process
directions,} the
independent variables of the algebraic problem: those are the
derivatives of the constitutive state, that are not included in the constitutive space. The choice of the constitutive state space is crucial
and can result in different kind of restricted constitutive
functions with Liu procedure.

As we have already mentioned, the weakly nonlocal constitutive spaces contain space derivatives.  In this cases some derivatives of the constraints - balances and others - can appear as additional constraints  that further restrict the entropy inequality. Whether these additional constraints should be considered or not depends on the order of weak nonlocality, the structure of the constraints and physical considerations. This is the  peculiarity of the exploitation of the Second Law for weakly nonlocal systems.    

In the following sections we will give several examples to demonstrate the application of the formalism.
\section{Thermodynamic evolution of internal variables}

In this section we investigate the thermodynamic restrictions on the evolution equation of internal variables in continua at rest. First in a first order weakly nonlocal constitutive state space, then in a second order weakly nonlocal one and finally considering the peculiarities of dual internal variables.  Further details of the related calculations and the physical interpretation can be found in 
\cite{Van05a,Van05a1,VanAta08a}. 

\subsection{First order nonlocality - relaxation}

Let us investigate the thermodynamic restrictions for the evolution of a classical internal variable field $a(t,\bf r)$, in a continuum at rest related to an inertial observer. There are no constraints or knowledge regarding the form of its evolution equation. Therefore the evolution equation can be given in a general form as
\begin{equation}
\underline{\partial_t a} + \hat{f} = 0,
\label{evi1}\end{equation}  
with an arbitrary constitutive function $\hat f$. Here and in the following the partial time derivatives are denoted by $\partial_t$ and the constitutive quantities are denoted by a hat $(\ \hat{}\ )$.  First we introduce a {\em first order weakly nonlocal constitutive state space} spanned by the basic fields $a$ and their gradients $\partial_i a$, where $i=1,2,3$. The notation with indices with Einstein summation convention is applied by distinguishing covariant and contravariant components (vectors and covectors) by upper and lower indices, e.g. $\partial_i J^i (= Div \mathbf{J}=\nabla\cdot \mathbf J)$ denotes the divergence of the vector field $J^i$. 

The above evolution equation is not completely arbitrary, it is restricted by the Second Law of thermodynamics. Therefore we assume that there is an entropy balance with a nonnegative  production term
\begin{equation}
\partial_t \hat s + \partial_i \hat J^i \geq 0.
\label{sbal}\end{equation}

Here the entropy density $\hat s$ and the entropy flux $\hat J^i$ are constitutive quantities, too. Therefore in this case 
\begin{itemize}
\item the basic state space is spanned by  $a$,
\item the constitutive state space is spanned by $(a,\partial_i a)$,
\item the constitutive functions are $\hat s, \hat J^i$ and $\hat f$. 
\end{itemize}
We may develop the derivatives according to the constitutive assumptions as 
\begin{equation}
\partial_a \hat s \underline{\partial_t a} +
  \partial_{\partial_ia}\hat s\; \underline{\partial_{it} a}+
     \partial_a\hat J^i\; \partial_i a + 
        \partial_{\partial_ja}\hat J^i\; \underline{\partial_{ij} a}\geq 0.
\label{sbal1}\end{equation}

Here the partial derivatives of the constitutive functions are denoted in an abbreviated manner, e.g. \(\frac{\partial \hat s}{\partial a}=\partial_a \hat s\). The underlined partial derivatives of the basic field are not in the constitutive space, they are independent algebraic quantities
and span the process direction space $(\partial_t a, \partial_{ti}a,\partial_{ij}a)$. Then we can apply the theorem of Liu identifying the underlined partial derivatives in (\ref{evi1}) and (\ref{sbal1}) by {\bf p} and their coefficients by {\bf a} and {\bf b} respectively. Now we introduce a Lagrange-Farkas multiplier $\hat \lambda$, which is a constitutive quantity, for the evolution equation (\ref{evi1}) and apply Liu procedure to determine the form of the constitutive functions, required by the entropy inequality:
\begin{equation}
  0  \leq \left(\partial_a \hat s - 
  \hat\lambda \right) \underline{\partial_t a} +
  \partial_{\partial_ia}\hat s\; \underline{\partial_{it} a} + 
        \partial_{\partial_ja}\hat J^i\; \underline{\partial_{ij} a}+
     \partial_a\hat J^i\; \partial_i a-
     \hat \lambda \hat f.    
\label{relaxliu}\end{equation}

Here the existence of the multiplier $\hat\lambda$ follows from Liu's theorem and we will  exploit that the process directions are independent of the constitutive space in the sense that the values of these derivatives can be different for same values of its coefficients depending e.g. on the initial conditions of (\ref{evi1}).
Therefore the multipliers of the underlined terms in (\ref{relaxliu}) are zero  and give the Liu equations:
\begin{eqnarray}
 \partial_t a &:& \partial_a \hat s =\hat \lambda, \\
  \partial_{it} a &:& \partial_{\partial_ia} \hat s =0^i, \\
   \partial_{ij} a &:& \partial_{\partial_ja} \hat J^i =0^{ij}. 
\end{eqnarray}

The first equation determines the Lagrange-Farkas multiplier and the last two ones
 show that both the entropy and the entropy flux is local, independent of the gradient of $a$. This is a complete solution of the system. The residual inequality is
\begin{equation}
0 \leq \partial_i \hat J^i(a)-
   \partial_a \hat s(a) \;\hat f(a,\partial_i a).
\end{equation}

Now we assume that the residual entropy flux is zero $\hat J^i \equiv 0^i$. This is the situation in isotropic materials according to the representation theorem of  isotropic vector functions. Therefore the entropy inequality reduces to a flux-force system. If we want to determine
the form of the evolution equation then the entropy should be considered as a given function and $\hat f$ is an undetermined constitutive quantity. The classical  solution of the above inequality is that the constitutive function $\hat f$ is proportional to the entropy derivative by a nonnegative constitutive multiplier:
\begin{equation}
\hat f = -\hat l\; \partial_a\hat s, \qquad \hat l>0.
\label{OCrel}\end{equation}

This is a general solution of the inequality if we assume two times differentiability of $\hat{f}$ (due to the mean value theorem of Lagrange, see e.g. \cite{Gur96a}). In the practice we rarely exploit this generality and restrict ourselves to the constant $\hat l$ case. This remark applies further in (\ref{ons2}), (\ref{evd1})-(\ref{evd2}) and  (\ref{ons3}). 

Therefore the  evolution equation of an internal variable restricted by the Second Law in isotropic materials is a relaxation type differential equation:
\begin{equation}
\partial_t a = \hat{l}\partial_a \hat s.
\label{fevi1}\end{equation}

Now let us summarize the most important steps of the procedure:

\begin{itemize}
\item Identification of the basic inequality and the basic constraints and the domain of the corresponding functions (constitutive state space). 
\item 
Performing the partial derivations and the identification of the process direction space by the partial derivatives of the basic state space and introducing the additional (derivative) constraints in case of necessity.
\item Application of Liu's theorem.
\item Solution of the Liu equations and the dissipation inequality. 
\end{itemize}  

In this example there was no need to apply additional derivative constraint, but it can be necessary incase of more extended constitutive state spaces as one can see in the next problem 

\subsection{Second order nonlocality - Ginzburg-Landau equation}

In this case we face to a similar problem, we are to determine the thermodynamic restrictions on the general evolution equation (\ref{evi1}), but now we assume that there is a second order weakly nonlocal state space spanned by the basic field $a$ and its first and second space derivatives $\partial_i a$ and $\partial_{ij} a$. 

Therefore in our second example 
\begin{itemize}
\item the basic state space is spanned by  $a$,
\item the constitutive state space is spanned by $(a,\partial_i a, \partial_{ij}a)$,
\item the constitutive functions are $\hat s, \hat J^i$ and $\hat f$. 
\end{itemize}

The corresponding process direction space is spanned by the next derivatives $(\partial_t a,\partial_{ti} a, \partial_{tij}a, \partial_{ijk}a)$. Let us observe that these are not independent any more, the gradient of (\ref{evi1}) is a linear   relation on the process direction space. Therefore we should consider 
\begin{equation}
\partial_{ti} a + \partial_i\hat{f} = 0_i
\label{devi1}\end{equation} 
as a further constraint to the entropy inequality (\ref{sbal}). We introduce  the Lagrange-Farkas multipliers $\hat\lambda$ for  (\ref{evi1}) and $\hat\Lambda^i$ for  (\ref{devi1}). Let us apply Liu procedure again, but in this case not separating the different steps:
\begin{eqnarray}
0 & \leq& \partial_t \hat s+ \partial_i \hat J^i-
   \hat \lambda\left(\partial_t a + \hat f\right)-
   \hat\Lambda^i\left(\partial_{ti} a + \partial_i\hat f\right)
 \nonumber\\
   &=&\partial_a\hat s\; \partial_t a + 
        \partial_{\partial_ia}\hat s\; \partial_{it} a+
     \partial_{\partial_{ij}a}\hat s\; \partial_{ijt} a +
      \partial_a\hat J^i\; \partial_i a + 
        \partial_{\partial_ja}\hat J^i\; \partial_{ij} a +
        \partial_{\partial_{jk}a}\hat J^i\; \partial_{ijk} a
    \nonumber\\
  &&-\ \hat \lambda \left( \partial_t a+ \hat f\right) -
      \hat \Lambda^i\left(\partial_{ti} a + 
      \partial_a\hat f\; \partial_i a + 
        \partial_{\partial_ja}\hat f\; \partial_{ij} a+
     \partial_{\partial_{jk}a}\hat f\; \partial_{ijk} a  \right)
     \nonumber\\
  &=& \left(\partial_a \hat s - 
        \hat\lambda \right) \underline{\partial_t a} +
  \left(\partial_{\partial_ia}\hat s - 
        \hat \Lambda^i \right)\; \underline{\partial_{it} a}+
    \partial_{\partial_{ij}a}\hat s\; \underline{\partial_{ijt} a} +
     \nonumber\\
  && \quad\left( \partial_{\partial_{jk}a}\hat J^i- 
         \hat \Lambda^i \partial_{\partial_{jk}a}\hat f 
                \right)\underline{\partial_{ijk} a}+
        \partial_a\hat J^i\; \partial_i a + 
        \partial_{\partial_ja}\hat J^i\; {\partial_{ij} a}-
    \nonumber\\
  && \quad\hat{\Lambda}^i\left( 
      \partial_a\hat f\; \partial_i a + 
        \partial_{\partial_ja}\hat f\; \partial_{ij} a\right)-\hat \lambda \hat f.    
\label{GLliu}\end{eqnarray}

We can see that the degenerations are different than in the previous case. In the following we do not give the detailed expositions of the theorem, but everybody can reconstruct that from the detailed presentation of the calculations.

The multipliers of the underlined partial derivatives in (\ref{GLliu}), the process direction space, give the Liu equations:
\begin{eqnarray}
 \partial_t a &:& \partial_a \hat s =\hat \lambda, \label{liugl1}\\
  \partial_{it} a &:& \partial_{\partial_ia} \hat s =\hat\Lambda^i, \\
   \partial_{ijt} a &:& \partial_{\partial_{ij}a} \hat s =0^{ij}, \\
  \partial_{ijk} a &:& \partial_{\partial_{jk}a}\hat J^i= 
         \hat \Lambda^i \partial_{\partial_{jk}a}\hat f . 
\label{liugl2}\end{eqnarray}

The first two equations determine the Lagrange-Farkas multipliers by the entropy derivatives and the third equation  shows that the entropy is independent of the second gradient of $a$. As a consequence, the Lagrange-Farkas multiplier $\hat\Lambda^i$ is independent of that derivative, therefore the last equation can be integrated and gives
\begin{equation}
\hat J^i (a,\partial_i a, \partial_{ij}a)= 
        \partial_{\partial_ia} \hat s(a,\partial_i a)\ 
        \hat f(a,\partial_i a, \partial_{ij}a) +
        \hat{\mathfrak{J}}^{i}(a,\partial_i a),
\end{equation} 
 where the residual entropy flux, $\hat {\mathfrak{J}}^i$, is an arbitrary constitutive function and the variables of the constitutive functions are explicitly written. This is a complete solution of the system of the Liu equations (\ref{liugl1})-(\ref{liugl2}). Therefore the  dissipation inequality simplifies to the following form
\begin{equation}
0 \leq \partial_i \hat{\mathfrak{J}}^i +
        \left(\partial_i(  \partial_{\partial_ia} \hat s)-
   \partial_a \hat s\right) \,\hat f.
\label{ons2}\end{equation}

Assuming that the residual entropy flux is zero $\hat {\mathfrak{J}}^i \equiv 0^i$, the entropy inequality reduces to a flux-force system. In this case the classical solution of the inequality is \begin{equation}
\hat f = \hat l\; \left(\partial_i(  \partial_{\partial_ia} \hat s)-
   \partial_a \hat s\right), \qquad \hat l>0.
\end{equation}

Therefore the form of the evolution equation of an internal variable in a second order weakly nonlocal constitutive state space is the Ginzburg-Landau equation:
\begin{equation}
\partial_t a = \hat l\; \left(\partial_a \hat s -
        \partial_i(  \partial_{\partial_ia} \hat s)\right).
\label{fevi11}\end{equation}  

We can get the classical form of the equation with a particular form of the entropy density \(\hat s(a,\partial_ia) = s_o(a)-\gamma(\partial_ia)^2\), with quadratic dependence on the gradient and with a constant coefficient \(\gamma\). $\gamma>0$ because of the concavity of the entropy. Other nonlocal thermodynamic potentials also can be used in the above derivation, e.g. the free energy, but one should be careful to use a correct thermodynamic structure when coupling to other thermodynamic interactions beyond the one related to the internal variable \cite{PenFif90a}. 

Ginzburg-Landau equation and its variants appear in different fields
of physics and are applied to several phenomena. Beyond their original appearance in superconductivity they play an important role in  pattern formation and they are the prototypical phase field models \cite{BroSpr96b,AraKra02a}. 

As we have mentioned in the introduction the traditional derivation of the Ginzburg-Landau equation has two
main ingredients:
\begin{itemize}
\item The static, equilibrium part is derived from a variational
principle.
\item The dynamic part is added by stability
arguments (relaxational form).
\end{itemize}

The physical content of the two ingredients of the classical derivation is sound and transparent
\cite{HohHal77a}. On the other hand,  the origin of the variational principle, the coupling of the two parts and the role of the Second Law of thermodynamics is ad-hoc and is
not compatible with the general balance and constitutive structure of continuum physics. Here we unified the ingredients  in a thermodynamic
derivation, and we derived the form of the entropy flux, too. We did not refer to any kind of variational
principle, however, the derived static part  has a complete
Euler-Lagrange form. The dynamic part contains a first order time
derivative, therefore one cannot hope to derive it from a
variational principle of Hamiltonian type \cite{VanMus95a}. The static part belongs to the zero entropy production, in this sense it is reversible. In our
approach we get the "reversible" part as a specific case of the
thermodynamic, irreversible thinking.

There are several alternate derivations based on different concepts \cite{Gur96a,Mar02a}. Here
we have demonstrated, that the physical assumptions to get the
Ginzburg-Landau equation are very moderate, the typical Ginzburg-Landau structure is a
straightforward consequence of the entropy inequality without any further additional set of concepts like the microforce balance or a variational principle.

\subsection{Dual internal variables - Hamiltonian structure}

 There are two classical approaches to determine the evolution of internal variables.

 When the evolution equations of internal variables are constructed exploiting
the entropy inequality, using exclusively thermodynamic
principles, then the  corresponding variables are called {\em internal
variables of state} \cite{Mau06a}. This frame has the advantage of
operating with familiar thermodynamic concepts (thermodynamic force,
entropy), however, no inertial effects are considered. The
thermodynamic theory of internal variables has a rich history (see
the historical notes in \cite{MauMus94a1}). A first more or less
complete thermodynamic theory was suggested by Coleman and Gurtin
\cite{ColGur67a}, and the clear presentation of the general ideas of
the theory was given by Muschik \cite{Mus90a1}. Internal variables
of state were applied for several phenomena in different areas of physics,
biology, and material sciences. A complete description of the
thermodynamic theory with plenty of applications based on this
concept of internal variables of state can be found in
\cite{Ver97b}.

There is a {second method} that generates the kinetic relations through the
Hamiltonian variational principle and suggests that inertial effects
are unavoidable. This approach has a mechanical flavor, and the
corresponding variables are called {\em dynamic degrees of
freedom}. Dissipation is added by dissipation potentials. This
theoretical frame has the advantage of operating with familiar
mechanical concepts (force, energy). The method was suggested by
Maugin \cite{Mau90a}, and it also has a large number of applications
\cite{Mau99b,Eng97b}. The clear distinction between these two
methods with a number of application areas is given by Maugin and
Muschik \cite{MauMus94a1,MauMus94a2} and Maugin \cite{Mau06a}.

Here we follow the terminology of Maugin and Muschik
\cite{MauMus94a1} with some important extensions. We call internal
variables of state those physical field quantities - beyond the
classical ones - whose evolution is determined by thermodynamical
principles. We call internal degrees of freedom those physical
quantities - beyond the classical ones - whose dynamics is
determined by mechanical principles.

One of the questions concerning this doubled theoretical frame is
related to common application of variational principles and
thermodynamics. Basic physical equations of thermodynamical origin
do not have variational formulations, at least without any further
ado \cite{VanMus95a}. That is well reflected by the appearance of
dissipation potentials as separate theoretical entities in
variational models dealing with dissipation.

On the other hand, with pure thermodynamical methods - in the
internal variables approach - inertial effects are not considered.
Therefore, the coupling to simplest mechanical processes seemingly
requires some additional assumptions, those are usually new
principles of mechanical origin.

On the other hand, with pure thermodynamic methods – in the internal variables
approach – inertial effects are not considered. Therefore, the coupling
to simplest mechanical processes seemingly requires introducing some improvements,
which are usually new principles of mechanical origin.

In the following we show that the mechanical structure arises from thermodynamic principles. Our suggestion requires dual internal
variables and a particular  generalization of the usual postulates of
non-equilibrium thermodynamics: we do not require  reciprocity relations. With dual internal
variables we are able to get inertial effects and to reproduce
the evolution of dynamic degrees of freedom. It could be impossible
with a single internal variable. This is the price we pay for the
generalization. In other words, instead of the doubling of the
theoretical structure we suggest the doubling of the number of
internal variables.

Let us consider a thermodynamic system where the state space is
spanned by two scalar internal variables $a$ and $b$. Then the
evolution of these variables is determined by the following
differential equations
 \begin{eqnarray} 
        \partial_t{a}+\hat f&=& 0, \label{d1}\\
        \partial_t{b}+\hat g&=& 0. \label{d2}
 \end{eqnarray}

The functions $\hat f$ and $\hat g$  are constitutive functions, restricted by the Second Law of thermodynamics. The entropy inequality, the main ingredient of the Second Law can be written in the same form as previously in (\ref{sbal}). The domain of the constitutive
functions (our constitutive space) is spanned by the state space
variables and by their first and second gradients. Therefore in this case
\begin{itemize}
\item the basic state space is spanned by  $(a,b)$,
\item the constitutive state space is spanned by $(a,\partial_i a, \partial_{ij}a, b, \partial_i b,\partial_{ij}b)$,
\item the constitutive functions are $\hat s, \hat J^i, \hat f$ and  $\hat g$. 
\end{itemize} This is a weakly
nonlocal constitutive space with second order weak nonlocality in
both variables.
The corresponding process direction space is spanned by the next derivatives $(\partial_t a,\partial_{ti} a, \partial_{tij}a, \partial_{ijk}a,\partial_t b,\partial_{ti} b, \partial_{tij}b, \partial_{ijk}b)$. The gradients of the evolution equations (\ref{d1})-(\ref{d2}) are constraints for the entropy inequality in the framework of a
second order constitutive state space for both of our variables 
\begin{eqnarray}
\partial_{ti} a &+& \partial_i\hat{f} = 0_i,\label{dd1}\\
\partial_{ti} b &+& \partial_i\hat{g} = 0_i.
\label{dd2}\end{eqnarray}
 
We introduce  the Lagrange-Farkas multipliers $\hat\lambda_{a}, \hat\lambda_{b}$ for  (\ref{d1})-(\ref{d2}) and $\hat\Lambda^i_a, \hat\Lambda^i_b$ for  
(\ref{dd1})-(\ref{dd2}), respectively. Liu procedure results in the Ginzburg-Landau structure of the previous subsection in a doubled form
\begin{eqnarray}
0 & \leq& \partial_t \hat s+ \partial_i \hat J^i-
   \hat \lambda_a\left(\partial_t a + \hat f\right)-
   \hat\Lambda_a^i\left(\partial_{ti} a + \partial_i\hat f\right)- 
\nonumber\\
   && \qquad\qquad\qquad\qquad\qquad
   \hat \lambda_b\left(\partial_t b + \hat g\right)-
   \hat\Lambda_b^i\left(\partial_{ti} b + \partial_i\hat g\right)
 \nonumber\\
   &=&\partial_a\hat s\; \partial_t a\! + 
        \partial_{\partial_ia}\hat s\;\partial_{it} a\!+
     \partial_{\partial_{ij}a}\hat s\; \partial_{ijt} a\!+
      \partial_a\hat J^i\; \partial_i a \!+ 
        \partial_{\partial_ja}\hat J^i\; \partial_{ij} a\! +
        \partial_{\partial_{jk}a}\hat J^i\; \partial_{ijk} a\!-
    \nonumber\\
  &&\quad  \hat \lambda_a \left( \partial_t a+ \hat f\right) -
      \hat \Lambda^i_a\left(\partial_{ti} a + 
      \partial_a\hat f\; \partial_i a + 
        \partial_{\partial_ja}\hat f\; \partial_{ij} a+
     \partial_{\partial_{jk}a}\hat f\; \partial_{ijk} a +\right. 
    \nonumber\\
  && \quad\left.  \partial_b\hat f\; \partial_i b + 
        \partial_{\partial_jb}\hat f\; \partial_{ij} b+
     \partial_{\partial_{jk}b}\hat f\; \partial_{ijk} b \right)
     \nonumber\\
       &+&\partial_b\hat s\; \partial_t b + 
        \partial_{\partial_ib}\hat s\; \partial_{it} b+
     \partial_{\partial_{ij}b}\hat s\; \partial_{ijt} b +
      \partial_b\hat J^i\; \partial_i b + 
        \partial_{\partial_jb}\hat J^i\; \partial_{ij}b +
        \partial_{\partial_{jk}b}\hat J^i\; \partial_{ijk}b-
    \nonumber\\
  &&\quad \hat \lambda_{b} \left( \partial_t b+ \hat g\right) -
      \hat \Lambda^i_b\left(\partial_{ti} b + 
      \partial_b\hat g\; \partial_i b + 
        \partial_{\partial_jb}\hat g\; \partial_{ij} b+
     \partial_{\partial_{jk}b}\hat g\; \partial_{ijk} b+\right. 
    \nonumber\\
  && \quad\left.  \partial_a\hat g\; \partial_i a + 
        \partial_{\partial_ja}\hat g\; \partial_{ij} a+
     \partial_{\partial_{jk}a}\hat g\; \partial_{ijk} a\right).
\end{eqnarray}

After some rearrangements one can get

\begin{eqnarray}
0 & \leq& 
    (\partial_a\hat s - \hat \lambda_a)\partial_t a + 
    (\partial_{\partial_ia}\hat s - \hat \Lambda^i_a)\partial_{it} a+
    \partial_{\partial_{ij}a}\hat s\;  \partial_{ijt} a +
 \nonumber\\
   &&\quad(\partial_{\partial_{jk}a}\hat J^i -
       \hat \Lambda^i_a \partial_{\partial_{jk}a}\hat f - 
       \hat \Lambda^i_b \partial_{\partial_{jk}a}\hat g )\partial_{ijk}a-
 \nonumber\\
    &&\quad  \hat \lambda_a \hat f -
       \hat \Lambda^i_a\left(\partial_a\hat f\; \partial_i a + 
       \partial_{\partial_ja}\hat f\; \partial_{ij} a+
       \partial_b\hat f\; \partial_i b + 
       \partial_{\partial_jb}\hat f\; \partial_{ij} b \right)+  
 \nonumber\\
    &&\quad
      \partial_a\hat J^i\; \partial_i a + 
      \partial_{\partial_ja}\hat J^i\; \partial_{ij} a 
 \nonumber\\
    &+&(\partial_b\hat s-\hat \lambda_{b})\partial_t b + 
       (\partial_{\partial_ib}\hat s -\hat \Lambda^i_b )\partial_{it} b+
       \partial_{\partial_{ij}b}\hat s\; \partial_{ijt} b +
 \nonumber\\ 
    &&\quad(\partial_{\partial_{jk}b}\hat J^i-
      \hat \Lambda^i_a\partial_{\partial_{jk}b}\hat f-
      \hat \Lambda^i_b\partial_{\partial_{jk}b}\hat g\;)\partial_{ijk}b-
 \nonumber\\
    &&\quad \hat \lambda_{b}\hat g -
      \hat \Lambda^i_b\left(\partial_b\hat g\; \partial_i b + 
        \partial_{\partial_jb}\hat g\; \partial_{ij} b+  
        \partial_a\hat g\; \partial_i a + 
        \partial_{\partial_ja}\hat g\; \partial_{ij} a\right)+
 \nonumber\\
    &&\quad
      \partial_b\hat J^i\; \partial_i b + 
      \partial_{\partial_jb}\hat J^i\; \partial_{ij}b .\nonumber
\end{eqnarray}

Here the multipliers of the process direction space give the Liu equations:
\begin{eqnarray}
 \partial_t a &:& \partial_a \hat s =\hat \lambda_a,  \label{liudu1}\\
  \partial_{it} a &:& \partial_{\partial_ia} \hat s =\hat\Lambda^i_{a}, \\
   \partial_t b &:& \partial_b \hat s =\hat \lambda_b, \\
  \partial_{it} b &:& \partial_{\partial_ib} \hat s =\hat\Lambda_b^i, \\
   \partial_{ijt} a &:& \partial_{\partial_{ij}a} \hat s =0^{ij}, \\
  \partial_{ijt} b &:& \partial_{\partial_{ij}b} \hat s =0^{ij}, \\
  \partial_{ijk} a &:& \partial_{\partial_{jk}a}\hat J^i= 
         \hat \Lambda_a^i \partial_{\partial_{jk}a}\hat f + 
         \hat \Lambda^i_b \partial_{\partial_{jk}a}\hat g,\\
   \partial_{ijk} b &:& \partial_{\partial_{jk}b}\hat J^i= 
         \hat \Lambda^i_b \partial_{\partial_{jk}b}\hat g+ 
         \hat \Lambda_a^i \partial_{\partial_{jk}b}\hat f.   \label{liudu2}
\label{liudu1v}\end{eqnarray}

The first four equations determine the Lagrange-Farkas multipliers by the entropy derivatives. The fifth and the sixth\ one show that the entropy is independent of the second gradient of $a$ and \(b\). Consequently, the Lagrange-Farkas multipliers $\hat\Lambda^i_{a}$ and $\hat\Lambda^i_{b}$ are independent of that derivatives, therefore the last two equations can be integrated and give
\begin{equation}
\hat J^i = 
        \partial_{\partial_ia} \hat s\ \hat f +
        \partial_{\partial_ib} \hat s\ \hat g +
        \hat{\mathfrak{J}}^{i}(a,\partial_i a, b,\partial_i b).
\label{ecd}\end{equation} 
 Here the variables of the the residual entropy flux $\hat {\mathfrak{J}}^i$, an arbitrary constitutive function, are explicitly written. This is a complete solution of the system of Liu equations (\ref{liudu1})-(\ref{liudu1v}). Therefore the dissipation inequality simplifies to the following form
\begin{equation}
0 \leq \partial_i \hat{\mathfrak{J}}^i +
        \left(\partial_i(  \partial_{\partial_ia} \hat s)-
   \partial_a \hat s\right)\hat f +
   \left(\partial_i(  \partial_{\partial_ib} \hat s)-
   \partial_b \hat s\right) \hat g.
\end{equation}

Now assuming that the residual entropy flux is zero, $\hat {\mathfrak{J}}^i \equiv 0^i$, the entropy inequality reduces to a flux-force system of the following form:

\begin{center}
\begin{tabular}{cccc}&&&\\
a-force:  & 
$\hat{A}$ =
        $\partial_i(\partial_{\partial_ia} \hat s)-\partial_a \hat s,$ &
\qquad a-flux: & $\hat{f}$, \\ & & &\\
b-force: & 
$\hat{B}$ = $\partial_i(\partial_{\partial_ib} \hat s)- \partial_b \hat s$, & 
\qquad b-flux: & $\hat g$. \\
&&&
\end{tabular}\end{center}

\ The classical solution of the entropy inequality gives a coupling of fluxes and forces as a system of coupled Ginzburg-Landau equations:
\begin{eqnarray}
\partial_t a =\hat f &=& \hat l_1 \left(\partial_i(  \partial_{\partial_ia} \hat s)-
        \partial_a \hat s\right)+
   \hat l_{12} \left(\partial_i(\partial_{\partial_ib} \hat s)-
   \partial_b \hat s\right),\label{evd1}\\
\partial_t a= \hat g &=& \hat l_{21} \left(\partial_i(  \partial_{\partial_ia} \hat s)-
        \partial_a \hat s\right)+
   \hat l_{2} \left(\partial_i(\partial_{\partial_ib} \hat s)-
   \partial_b \hat s\right).   
\label{evd2}\end{eqnarray}

The constitutive Onsagerian coefficients  $\hat l_{1}, \hat l_{2}, \hat l_{12}, \hat l_{21}$ are  restricted by the Second Law. For the sake of generality we do not assume any kind of reciprocity here. We decouple the symmetric and antisymmetric parts introducing $\hat l=(\hat l_{12}+\hat l_{21})/2$ and $\hat k=(\hat l_{12}-\hat l_{21})/2$. The entropy production is nonnegative if
\begin{gather}
\hat l_{1}>0, \quad \hat l_{2}>0 \quad \text{and} \quad \hat l_{1} \hat l_{2} -\hat l^2 \geq 0
\end{gather}
Now the evolution equations (\ref{evd1})-(\ref{evd2}) can be written equivalently  as  
\begin{eqnarray}
 \partial_t a &=&  \hat k \hat B + \hat l_1 \hat A+ \hat l \hat B, \label{o1} \\
 \partial_t b &=& -\hat k \hat A  + \hat l \hat A + \hat l_2 \hat B
\label{o2}.\end{eqnarray}

\subsubsection{Remark on dissipation potentials}

We may introduce dissipation potentials for the dissipative part of
the equations, if the condition of their existence is satisfied.
Dissipation potentials are the children of variational principles, they are artificially added to a set of reversible equations of variational origin to generate some kind of dissipative effects. In our case there is no need of this assumption, our construction gives the most general dissipative system without any further ado. 
Moreover, here it is clear what belongs to the dissipative part and
what belongs to the nondissipative part of the  evolution
equations. The terms with the symmetric conductivity contribute to
the entropy production and the terms from the skew symmetric part do
not. On the other hand,  there is no need of potential construction,
as we are not looking for a variational formulation. Moreover,  the
symmetry relations are not sufficient for the existence of
dissipation potentials in general, as we have emphasized previously.
In  case of constant coefficients (strict linearity), the
dissipation potentials always exist for the dissipative (symmetric)
part.
 
\subsubsection{Remark on the reciprocity relations}

The reciprocity relations are the main results of the great idea of
Lars Onsager connecting fluctuation theory to macroscopic
thermodynamics \cite{Ons31a1,Ons31a2,OnsMac53a,MacOns53a,Cas45a}. As
it was written by Onsager himself on the validity of his result:
"The restriction was stated: on a kinetic model, the thermodynamic
variables must be algebraic sums of (a large number of) molecular
variables, and must be {\em even} functions of those molecular
variables which are odd functions of time (like molecular
velocities)" \cite{MacOns53a}. The Casimir reciprocity relations are
based on microscopic fluctuations, too \cite{Cas45a}. We do not have
such a microscopic background for most of internal variables. E. g.
in the case of damage the internal variables are
reflecting a structural disorder on a mesoscopic scale. The relation
between thermodynamic variables and the microscopic structure is
hopelessly complicated. On the other hand, the Onsagerian
reciprocity is based on time reversal properties of corresponding
physical quantities either at the macro or at the micro level.
Looking for the form of evolution equations without a microscopic
model, we do not have any information on the time reversal
properties of our physical quantities  neither at the micro- nor at
the macroscopic level. Therefore, we can conclude that lacking the
conditions of the Onsagerian or Casimirian reciprocity gives no
reasons to assume their validity in the internal variable theory.

Let us observe the correspondence of evolution equations for
internal variables with the reciprocity relations by means of a few
simple examples.

\subsubsection{Example 1: Internal variables}

Let us consider materials with diagonal
conductivity matrix ${\bf L}$  ($\hat l = 0$, $\hat k=0$). It is clear that
the Onsagerian reciprocity relations are satisfied, and we return to
the classical situation with fully uncoupled internal variables:
\begin{eqnarray*}
\partial_t a &=&   \hat l_1 \hat A , \\
\partial_t b &=&   \hat l_2 \hat B.
\end{eqnarray*}
In this case the evolution equations for dual internal variables
$a$ and $b$ are the same as in the case of single internal
variable.
 
\subsubsection{Example 2: Dynamic degrees of freedom}
\label{idf}

We now assume that all conductivity coefficients are constant, and
their values are $l_1 = l = 0$, $k=1$. This means that
$l_{12}=-l_{21}$, i.e., the Casimirian reciprocity relations are
satisfied.

For simplicity, we consider a specific decomposition of the entropy density into two parts,
which depend on different internal variables
\begin{displaymath}
\hat s(a,\partial_ia,b,\partial_ib) = -K(b) -
W(a, \partial_ia ).
 \label{p}
\end{displaymath}

The negative signs are introduced taking into account the concavity
of the entropy. Then the thermodynamic forces are represented as
$$
 \hat A = \partial_a W - \partial_i (\partial_{\partial_ia} W), \qquad
 \hat B = d_b K = K'(b),
 $$
and Eqs. (\ref{o1})-(\ref{o2}) are simplified to
\begin{eqnarray}
\partial_t a  &=&  \hat B = K'(b)  \label{so1} \\
\partial_tb  &=&  -\hat A + l_2 \hat B =
    -\partial_a W +
   \partial_i (\partial_{\partial_ia} W)+ l_2K'(b)
    \label{so2}.
\end{eqnarray} 
One may recognize that the obtained system of equation corresponds
exactly to  a Hamiltonian
form with the last term of (\ref{so2}) as added dissipation.  Here the entropy density $\hat s$ plays the role of a Hamiltonian density, $a$ is a Lagrangian variable and $b$ corresponds to the (slightly generalized) conjugated moment. The transformation into a Lagrangian form is trivial if $K$ is quadratic $K(b) = \frac{b^2}{2 m}$, where $m$ is a
constant. Then the whole system corresponds to the general structure of  dynamic degrees of freedom  as one can compare to Eq. (5.14) in \cite{MauMus94a1}  with the
Lagrangian
 $$
  \mathcal{L}(\partial_ta, a,\partial_i a) =
    m\frac{(\partial_ta)^2}{2} - W(a,\partial_ia),
  \quad \text{and} \quad
  D(a, \partial_ia) =  \frac{ml_2}{2}(\partial_ta)^2
$$
as dissipation potential. Moreover, the entropy flux density (\ref{ecd}) in case of our
special conditions can be written as
\begin{displaymath}
 {J}^i=
    -\partial_{\partial_ia}sK'(b)
    +\mathfrak{J}^{i}_0,
\end{displaymath} 
and one can infer that natural boundary conditions of the
variational principle related to the above Lagrangian 
correspond to the condition of
vanishing entropy flow at the boundary, with  $\mathfrak{J}^{i}_0\equiv0^i$.

Therefore, the variational structure of internal degrees of freedom
is recovered in the pure thermodynamic framework. The thermodynamic
structure resulted in several sign restrictions of the coefficients,
and the form of the entropy flux is also recovered. The natural
boundary conditions correspond to a vanishing extra entropy flux.

\subsubsection{Example 3: Diffusive internal variables \cite{MauDro83a,DroMau01a,EngVen00a}}

Now we give an additional example to see clearly the reduction of
evolution equations of internal degrees of freedom to evolution
equations for internal variables and the extension of the later to
the previous one.

We keep the values of conductivity coefficients (i.e., $l_1 = l =
0$, $k=1$), but assume that both $K$ and $W$ are quadratic functions
$$
K(b)=\frac{\beta}{2}b^{2}, \quad 
W(a,\partial_i a)=\frac{\alpha}{2}(\partial_ia)^{2},
$$
where $\alpha$ and $\beta$ are positive constants according to the concavity
requirement.

In this case, the evolution equations (\ref{so1})-(\ref{so2})
reduce to
\begin{eqnarray}
 \partial_ta &=& \hat B = K'(b)=\beta b,  \label{so3} \\
 \partial_tb &=& 
    - \partial_a W +\partial_i (\partial_{\partial_ia} W)+
    l_2K'(b)=\alpha\partial_{ii}a +l_2\beta b
    \label{so4}.
\end{eqnarray}
Putting $b$ from Eq. (\ref{so3}) into Eq. (\ref{so4}),
we have
\begin{equation}
  \frac{1}{\alpha\beta}\partial_{tt}a - \frac{l_2}{\alpha}\partial_ta=
  \partial_{ii} a. 
\label{so5}\end{equation}
which is a Cattaneo-Vernotte type hyperbolic equation (telegraph
equation) for the internal variable $a$. This can be considered
as an extension of a diffusion equation by an inertial term or as
and extension of a damped Newtonian equation (without forces) by a
diffusion term.

\subsubsection{Discussion}

In the framework of the thermodynamic theory with dual weakly
nonlocal internal variables we are able to
recover the evolution equations for internal degrees of freedom.

We have seen that the form of evolution equations depend on the
mutual interrelations between the two internal variables. In the
special case of internal degrees of freedom, the evolution of one
variable is driven by the second one, and vice versa. This can be
viewed as a duality between the two internal variables. In the case
of pure internal variables of state, this duality is replaced by
self-driven evolution for each internal variable. The general case
includes all intermediate situations.

It is generally accepted that internal variables are "measurable but
not controllable" (see e.g. \cite{Kes93a1}). Controllability can be
achieved by boundary conditions or fields directly acting on the
physical quantities. We have seen how natural boundary conditions
arise considering nonlocality of the interactions through weakly
nonlocal constitutive state spaces.

As we wanted to focus on generic inertial effects, our treatment is
simplified from several points of view. Vectorial and tensorial
internal variables were not considered and the couplings to
traditional continuum fields result in degeneracies and more
complicated situations than in our simple examples.

It is important to remark, that skew symmetric couplings are not
always related directly to inertial effects and indicate two
directions, one into mechanics and one into thermodynamics, where
our method can be generalized. Let us mention here the related
pioneering works of Verh\'as, where skew symmetric conductivity
equations appear in different inspiring contexts
\cite{Ver89a,Ver01p}.

Finally, let us mention that the idea of constructing a unified
theoretical frame for reversible and irreversible dynamics has a
long tradition. The corresponding research was not restricted to the
case of internal variables and was looking for a classical
Hamiltonian or a generalized variational principle that would be
valid for both dissipative and nondissipative evolution equations
(see, e.g., \cite{Gya70b,GlaPri71b,MarGam91a,VanNyi99a} and the
references therein).

\section{Classical Irreversible Thermodynamics}

In this section we investigate the evolution of extensive physical field quantities whose equation of motion is a balance and we will see that for first order weakly nonlocal  state spaces the thermodynamic force-flux system is a consequence  of the nonnegative entropy production. Treatment of Extended Irreversible Thermodynamics and  further details can be found in \cite{Van03a}.

 A balance type evolution equation for a conserved quantity {\bf a} can be given in a general form as
\begin{equation}
\partial_t {\bf a} + \partial_i \hat{\bf j}^i = 0.
\label{evbal}\end{equation}  

Here $i\in\{1,2,3\}$ denotes the spatial coordinates. The conserved quantity 
can be a Descartes product of densities of extensives of any tensorial order
  e.g. ${\bf a} = (\rho,e,p^j, ...)$, where $\rho$ is the mass density, $e$ is 
the internal energy density, $p^j$ is the momentum density. Our continuum is at rest, 
$\hat{\bf j}^i$ are the corresponding fluxes, e.g.  ${\bf j}^i = 
(\rho v^i, \rho e v^i+j_e^i,\rho v^jv^i+P^{ji}, ...)$, where the first term is 
the current of the mass, the second is the convective and conductive current 
of internal energy and the third is the current of momentum. Therefore with this convenient  notation we introduce only those free indices that are important from the point of view of the balance structure of the evolution equation.    For the balances of particular real physical quantities there are peculiarities that we do not introduce here. For example for a continuum at rest traditionally there is no mass flux as a consequence of barycentric velocities, the source terms can play an important role (for internal energy, chemical production, etc.). Those peculiarities introduce further constraints and additional terms in the final equations, that we do not consider in this general calculation to show the core structure of classical irreversible thermodynamics. Moreover in this case we restrict ourselves for a continuum at rest. Some consequences of the motion of the continua is considered and investigated at the next section. We assume here  a first order weakly nonlocal state space.
  
Therefore in this case 
\begin{itemize}
\item the basic state space is spanned by  ${\bf a}$,
\item the constitutive state space is spanned by $({\bf a},\partial_i {\bf a})$,
\item the constitutive functions are $\hat s, \hat J^i$ and $ \hat {\bf j}^i$. \end{itemize}

The above evolution equation is not completely arbitrary, but restricted by the Second Law of thermodynamics (\ref{sbal}). Let us introduce a 
Lagrange-Farkas multiplier $\lambda$ for the evolution equation (\ref{evbal}) and apply Liu procedure to determine the form of the constitutive functions, required by the entropy inequality:
\begin{eqnarray}
  0 & \leq& \partial_t \hat  s({\bf a},\partial_i {\bf a})+ \partial_i \hat J^i({\bf a},\partial_i {\bf a})-
   \hat \lambda\left(\partial_t {\bf a} + \partial_i \hat{\bf j}^i ({\bf a},\partial_i {\bf a})\right)
   \nonumber\\
  &=&\partial_{\bf a}\hat s\; \partial_t {\bf a} + 
        \partial_{\partial_i{\bf a}}\hat s\; \partial_{it} {\bf a}+
     \partial_{\bf a}\hat J^i\; \partial_i {\bf a} + 
        \partial_{\partial_j{\bf a}}\hat J^i\; \partial_{ij} {\bf a}
     \nonumber\\
  &&\qquad- \hat{\lambda} \left( \partial_t {\bf a}+ 
        \partial_{\bf a}\hat{\bf j}^i
         \partial_i {\bf a} + \partial_{\partial_j{\bf a}}\hat{\bf j}^i
                \partial_{ij} {\bf a}\right) 
     \nonumber\\
  &=& \left(\partial_{\bf a} \hat s - 
  \hat\lambda \right) \underline{\partial_t {\bf a}} +
  \partial_{\partial_i{\bf a}}\hat s\; \underline{\partial_{it} {\bf a}}+
     \left(\partial_{\bf a}\hat J^i -
       \hat{\lambda}\partial_{\bf a}\hat{\bf j} \right) \partial_i {\bf a}
   \nonumber\\
        &&\qquad + \left(\partial_{\partial_j{\bf a}}\hat J^i - 
       \hat{\lambda}\partial_{\partial_j{\bf a}}\hat{\bf j}^i  
       \right) \underline{\partial_{ij} {\bf a}}.
\end{eqnarray}

The underlined partial derivatives of the basic field are not in the constitutive space, they are independent algebraic quantities
and span the process direction space. The multipliers of those terms are zero according to Liu's theorem and give the Liu equations:
\begin{eqnarray}
 \partial_t {\bf a} &:& \partial_{\bf a} \hat s =\hat \lambda, \label{citliu1}\\
  \partial_{it} {\bf a} &:& \partial_{\partial_i{\bf a}} \hat s =0^i, \\
   \partial_{ij} {\bf a} &:& \partial_{\partial_j{\bf a}} \hat J^i =
        \hat\lambda\partial_{\partial_j{\bf a}}\hat{\bf j}^i. 
\label{citliu3}\end{eqnarray}

The first equation determines the Lagrange-Farkas multiplier and the second one  shows that the entropy should be local, independent of the gradient of $\bf  a$. Therefore one can solve the third equation as
\begin{equation}
\hat J^i({\bf a},\partial_i{\bf a}) = 
        \partial_{\bf a}s({\bf a})\cdot 
                \hat{\bf j}^i({\bf a},\partial_i{\bf a})+
            \hat{\mathfrak{J}}^i({\bf a}),
\label{citsj}\end{equation}
where we have denoted the variables of the constitutive functions and we have introduced the local residual entropy flux $\hat{\mathfrak{J}}^i$. 
Several authors suggest this kind of additive supplement to the classical entropy flux (e.g. the {\bf K} vector of M\"uller \cite{Mul68a}). This is a complete solution of the system (\ref{citliu1})-(\ref{citliu3}). The dissipation inequality is
\begin{equation}
0 \leq \partial_i \hat{\mathfrak{J}}^i+
   \hat{\bf j}^i\partial_i(\partial_{\bf a} \hat s).
\end{equation}

Assuming that the residual entropy flux is zero $\hat{\mathfrak J}^i \equiv 0^i$, the entropy inequality reduces to the usual flux-force system of Classical Irreversible Thermodynamics, where the thermodynamic forces are the gradients of the intensives and the thermodynamic fluxes are identical to the the fluxes of the extensives from the balances.
\begin{center}\begin{tabular}{cccc}
        Flux: & $\hat{\bf j}^i$, & \qquad
        Force: & $\partial_i(\partial_{\bf a} \hat s)$.
\end{tabular}
\end{center}

The classical  solution of the above inequality is that the fluxes are  proportional to the forces:
\begin{equation}
\hat{\bf j}^i =\hat {\bf L}^{ik}\partial_k(\partial_{\bf a} \hat s),
\label{ons3}\end{equation}
where $\hat{\bf L}^{ik}$ is symmetric and positive definite. Therefore balances of extensives are reduced to transport  equations of the form:
\begin{equation}
\partial_t {\bf a} + 
        \partial_i \left({\bf L}^{ij}\partial_j(\partial_{\bf a} \hat s)\right) = 0.
\label{fevbal}\end{equation}  

We have seen that in our calculations the classical form  of the
entropy production (the first term in the inequality above) {\em and}
the classical form of the entropy flux were consequences of the
Second Law, with the assumption of first order nonlocality. A rigorous
treatment showed that the seemingly intuitive steps of irreversible thermodynamic
modelling are well supported and explained by Liu procedure. We can see that 
the second part of the usual phenomenological formulation of the local 
equilibrium hypothesis \cite{GlaPri71b} is a consequence of the assumption of 
first order nonlocality: the state functions are those in equilibrium due to 
the local entropy.  

\ \ \section{One component fluids - second order nonlocal in the density}

 Up to know we have investigated evolution equations of internal variables and the general structure of Classical Irreversible Thermodynamics in a somewhat abstract manner. In this section we analyze a more particular and less abstract example where the physical meaning of the variables in the basic state space is well known. Our example is a one component heat conducting fluid where the constitutive state space for the energy and the velocity fields are first order, but second order in the density. We have seen that in case of first order weakly nonlocal state spaces   our method gives the well known classical structure with a somewhat simplified manner, where the number of independent assumptions are reduced. E.g. the form of the entropy flux is calculated and  not postulated, as we have demonstrated in the previous section. However, the second order nonlocality in the density variables leads to a surprising result and gives the viable family of Korteweg fluids, those that are compatible with the Second Law.  One can find more details in \cite{VanFul06a,VanFul04a}  and with alternate methods in  \cite{DunSer85a,Mor08a}.

\subsection{Fluid mechanics in general}

The {\em basic state space} of one-component fluid mechanics is
spanned by the mass density $\rho,$  the velocity ${v}^i$ and the energy density $e$ of the
fluid. Hydrodynamics is based on the balance of mass, energy and momentum \cite{Gya70b}. In classical fluid mechanics the constitutive space, the domain of the
constitutive functions, is spanned by the basic state space
$(\rho,v^i,e)$ and the gradient of the velocity $\partial_i v^j$ and the temperature $\partial_iT$. The pressure/stress tensor and the flux of the internal energy are the constitutive quantities
in the theory. 

The balance of mass can be written as
\begin{equation}
  \partial_t{\rho}+\partial_i(\rho v^i) = 0.
\label{cond_mass}\end{equation}

There is only a convective flux  for mass. The balance of momentum, i.e. the Cauchy equation, is
\begin{equation}
\partial_t(\rho v^i)+ \partial_j\left(\hat{P}^{ij} + 
        \rho v^iv^j\right) = 0^i,
\label{cond_mom}\end{equation}

\noindent where $\rho v^i$ is the momentum density and $\hat {P}^{ij}$ is the pressure tensor, the conductive flux of the momentum. 

The balance of energy is given as
\begin{equation}
  \partial_t{e}+\partial_i(\hat q^i+e v^i) = 0,
\label{cond_e}\end{equation}

Here $e$ is the total energy density and $\hat q^i$ is the conductive flux of the energy. The Second Law requires that the production of the entropy is
nonnegative in insulated and source-free systems, therefore source terms are not considered in the balances.  Hence in
our case there is no mass production, the external forces are absent and the energy is conserved. We do not need the concept of the internal energy yet, first we analyse the consequences of the Second Law with the balance of the total energy. As we are dealing with a fluid, it is convenient to separate the conductive and convective entropy fluxes. Hence  (\ref{sbal}) will be written as 
\begin{equation}
\partial_t \hat s + \partial_i (\hat J^i + \hat sv^i)\geq 0.
\label{csbal}\end{equation}

Therefore in this case
\begin{itemize}
\item the basic state space is spanned by  $(\rho, v^i, e)$,
\item the constitutive state space is spanned by 
$(\rho,\partial_i\rho, \partial_{ij}\rho, v^i, \partial_i v^j, e, \partial_ie)$,
\item the constitutive functions are $\hat s, \hat J^i, \hat {q}^i$ and $\hat{P}^{ij}$. \end{itemize}

It is somewhat convenient to introduce the relative velocity $v^i$ as basic state variable (instead of the momentum) and the gradient of the energy as constitutive state variable (instead of the temperature  gradient). As we have a second order weakly nonlocal extension in the mass density we need the gradient of (\ref{cond_mass}) as a further constraint in the entropy inequality
\begin{equation}
  \partial_{it}{\rho}+\partial_{ij}(\rho v^j) = 0_{i},
\label{cond_gmass}\end{equation}

We introduce the Lagrange-Farkas multipliers 
$\hat{\lambda},\hat{\Lambda}^i,\hat{\Gamma}_i,\hat{\gamma}$ for the balances (\ref{cond_mass}), (\ref{cond_gmass}), (\ref{cond_mom}) and (\ref{cond_e}) respectively.

Now we apply Liu procedure, with the method of Lagrange-Farkas
multipliers as
in the previous sections
\begin{eqnarray}
  0 & \leq& \partial_t \hat s+ \partial_i \hat J^i + \partial_i(\hat s v^i)-
   \hat \lambda\left( \partial_t{\rho}+\partial_i(\rho v^i)\right)-
     \hat{\Lambda}^i\left( \partial_{it}{\rho}+\partial_{ij}(\rho v^j)\right)-
      \nonumber\\
  &&\hat{\Gamma}_i\left(\partial_t(\rho v^i)+ 
        \partial_j\left(\hat{P}^{ij} + \rho v^iv^j\right)  \right) -
     \hat{\gamma}\left( \partial_t{e}+\partial_i(\hat q^i+e v^i)\right).
\label{sfliu1}\end{eqnarray}

Developing the partial derivatives of the constitutive functions gives:
\begin{gather*}
  \partial_\rho\hat{s} \partial_t\rho + 
  \partial_{\partial_i\rho}\hat{s} \partial_{ti}\rho+
  \partial_{\partial_{ij}\rho}\hat{s} \partial_{tij}\rho+
  \partial_{v^i}\hat{s} \partial_{t}v^i+
  \partial_{\partial_iv^j}\hat{s} \partial_{tj}v^i+
  \partial_e\hat{s} \partial_te + 
\\\partial_{\partial_ie}\hat{s} \partial_{ti}e+
 \partial_\rho\hat{J}^i \partial_i\rho + 
  \partial_{\partial_j\rho}\hat{J}^i \partial_{ij}\rho+
  \partial_{\partial_{jk}\rho}\hat{J}^i \partial_{ijk}\rho+
  \partial_{v^j}\hat{J}^i \partial_{i}v^j+
  \partial_{\partial_kv^j}\hat{J}^i \partial_{jk}v^i+
\\\partial_e\hat{J}^i \partial_ie + 
  \partial_{\partial_je}\hat{J}^i \partial_{ij}e+
\\\partial_i(\hat s v^i)-
  \hat\lambda\left( 
        \partial_t{\rho}+
        \rho\partial_i v^i+ 
        v^i\partial_i\rho\right)-
\\\hat{\Lambda}^i\left( 
        \partial_{it}{\rho}+
        \rho \partial_{ij}v^j+
        v^j\partial_{ij}\rho+
        \partial_{i}\rho\partial_j v^j+
        \partial_{j}\rho\partial_i v^j\right)-
\\\hat{\Gamma}_i\left(
    \rho\partial_t v^i+  v^i\partial_t\rho+ 
    \rho v^i\partial_jv^j+
    \rho v^j\partial_jv^i+
    v^j v^i\partial_j\rho+ 
  \partial_{\rho}\hat{P}^{ij}\partial_{j}\rho+
  \partial_{\partial_k\rho}\hat{P}^{ij}\partial_{jk}\rho+\right.
\\\left. 
 \partial_{\partial_{kl}\rho}\hat{P}^{ij}\partial_{jkl}\rho+ 
  \partial_{v^k}\hat{P}^{ij}\partial_{j}v^k+
  \partial_{\partial_{l}v^k}\hat{P}^{ij}\partial_{jl}v^k +
  \partial_{e}\hat{P}^{ij}\partial_{j}e+
  \partial_{\partial_ke}\hat{P}^{ij}\partial_{jk}e \right) -
\\\hat{\gamma}\left( 
  \partial_t{e}+ e\partial_i v^i+v^i\partial_i e+
  \partial_\rho\hat{q}^i \partial_i\rho + 
  \partial_{\partial_j\rho}\hat{q}^i \partial_{ji}\rho+\right.
\\\left.
  \partial_{\partial_{jk}\rho}\hat{q}^i \partial_{ijk}\rho+
  \partial_{v^j}\hat{q}^i \partial_{i}v^j+
  \partial_{\partial_kv^j}\hat{q}^i \partial_{ik}v^j+
  \partial_e\hat{q}^i \partial_ie + 
  \partial_{\partial_je}\hat{q}^i \partial_{ji}e\right)\geq 0.
\end{gather*}

A little rearrangement of the terms results in 
\begin{gather*}
  (\partial_\rho\hat{s}-
        \hat\lambda-
        \hat{\Gamma}_i v^{i})\partial_t\rho + 
  (\partial_{\partial_i\rho}\hat{s}-
    \hat{\Lambda}^i )\partial_{ti}\rho+
  \partial_{\partial_{ij}\rho}\hat{s} \partial_{tij}\rho+
  (\partial_{v^i}\hat{s} -
     \rho\hat{\Gamma}_i)\partial_{t}v^i+
\\\partial_{\partial_iv^j}\hat{s} \partial_{tj}v^i+
  (\partial_e\hat{s}-
    \hat\gamma) \partial_te + 
  \partial_{\partial_ie}\hat{s} \partial_{ti}e+
\\(\partial_{\partial_{jk}\rho}\hat{J}^i -
    \hat\Gamma_{l}\partial_{\partial_{jk}\rho}\hat{P}^{li}-
    \partial_{\partial_{jk}\rho}\hat{q}^i)\partial_{ijk}\rho+
\\(\partial_{\partial_kv^j}\hat{J}^i -
    \hat\Gamma_{l}\partial_{\partial_{k}v^i}\hat{P}^{lj}-
    \hat\gamma\partial_{\partial_kv^i}\hat{q}^j \partial_{ik}v^j 
   )\partial_{jk}v^i-
   \hat\Lambda^i \rho \partial_{ij}v^j+
\\(\partial_{\partial_je}\hat{J}^i -\hat\Gamma_l
  \partial_{\partial_ie}\hat{P}^{lj}+
  \hat\gamma\partial_{\partial_je}\hat{q}^i)\partial_{ij}e+ 
  \\\partial_\rho\hat{J}^i \partial_i\rho + 
  \partial_{\partial_j\rho}\hat{J}^i \partial_{ij}\rho+
  \partial_{v^j}\hat{J}^i \partial_{i}v^j+
  \partial_e\hat{J}^i \partial_ie + 
\\\partial_i(\hat s v^i)-
  \hat\lambda\left( 
        \rho\partial_i v^i+ 
        v^i\partial_i\rho\right)-
\\\hat{\Lambda}^i\left( 
        \rho \partial_{ij}v^j+
        v^j\partial_{ij}\rho+
        \partial_{i}\rho\partial_j v^j+
        \partial_{j}\rho\partial_i v^j\right)-
\\\hat{\Gamma}_i\left(
    \rho v^i\partial_jv^j+
    \rho v^j\partial_jv^i+
    v^j v^i\partial_j\rho+ 
  \partial_{\rho}\hat{P}^{ij}\partial_{j}\rho+
  \partial_{\partial_k\rho}\hat{P}^{ij}\partial_{jk}\rho+\right.
\\\left. 
 \partial_{v^k}\hat{P}^{ij}\partial_{j}v^k+
  \partial_{e}\hat{P}^{ij}\partial_{j}e \right) -
\\\hat{\gamma}\left( 
  e\partial_i v^i+v^i\partial_i e+
  \partial_\rho\hat{q}^i \partial_i\rho + 
  \partial_{\partial_j\rho}\hat{q}^i \partial_{ji}\rho+
  \partial_{v^j}\hat{q}^i \partial_{i}v^j+
  \partial_e\hat{q}^i \partial_ie\right)\geq 0.
\end{gather*}

Then the Liu-equations follow as the multipliers of the members of the process direction space, the derivatives that are out of the constitutive space
\begin{eqnarray}
  \partial_t \rho \ &:& \partial_\rho \hat{s}-\hat\lambda-
        \hat\Gamma_iv^i =0,\label{fliu1}\\
  \partial_{ti} \rho &:& \partial_{\partial_{i}\rho} \hat s-
        \hat\Lambda^i=0^i,  \label{fliu2}\\
  \partial_{tij} \rho &:& \partial_{\partial_{ij}\rho} \hat s = 0^{ij},
        \label{fliu3} \\
  \partial_t v^i \ &:& \partial_{v^i} \hat{s}-\rho\hat\Gamma_i =0_i,\label{fliu4}\\
  \partial_{tj} v^i &:& \partial_{\partial_{j}v^i} \hat s=0^j_i,  
        \label{fliu5}\\ 
  \partial_t e \ &:& \partial_e \hat{s}-\hat\gamma =0,\label{fliu6}\\
  \partial_{ti}e  &:& \partial_{\partial_{i}e} \hat s=0^i,  \label{fliu7}\\
  \partial_{ijk} \rho &:& \partial_{\partial_{kj}\rho} \hat J^i =
        \hat\Gamma_l\partial_{\partial_{kj}\rho}\hat{P}^{li}+
        \hat\gamma \partial_{\partial_{kj}\rho}\hat{q}^i,\label{fliu8}\\
  \partial_{jk}v^i  &:& \partial_{\partial_{k}v^i} \hat J^j =
        \hat\Gamma_l\partial_{\partial_{k}v^i}\hat{P}^{lj}+
        \hat\gamma \partial_{\partial_{k}v^i}\hat{q}_j+
        \frac{\rho}{2}\hat\Lambda^l(\delta_l^j\delta_i^k-\delta_l^k\delta_i^j),
        \label{fliu9}\\
  \partial_{ij} e &:& \partial_{\partial_{j}e} \hat J^i =
        \hat\Gamma_l\partial_{\partial_{i}e}\hat{P}^{lj}+
        \hat\gamma \partial_{\partial_{j}e}\hat{q}^i.\label{fliu10} 
\end{eqnarray}

As a consequence of (\ref{fliu3}), (\ref{fliu5}) and (\ref{fliu7}), the entropy density does not depend on $\partial_{ij}\rho$, $\partial_{j}v^i$ and 
$\partial_{i}e$. (\ref{fliu1}), (\ref{fliu2}), (\ref{fliu4}) and (\ref{fliu6}) give the Lagrange-Farkas multipliers in terms of the entropy derivatives.  Therefore, from a thermodynamic point
of view, the Lagrange-Farkas  multipliers are the  normal and generalized intensive variables \cite{KirHut02p}.  Now, one can give a solution
of (\ref{fliu8})-(\ref{fliu10}) as
\begin{eqnarray}
\hat{J}^i &=& \partial_e\hat{s}\hat{q}^i+
  \frac{\rho}{2}\left(\partial_{\partial_i\rho}\hat s \partial_jv^j+
        \partial_{\partial_j\rho}\hat s \partial_jv^i\right) +
  \frac{1}{\rho}\partial_{v^j}\hat s \hat{P}^{ji} + 
  \mathfrak{\hat J}^i(\rho,\partial_i\rho,v^i,e),
\end{eqnarray}

\noindent where the residual entropy flux $\mathfrak{\hat J}^i$ is an arbitrary function. Thus Liu's equations can be solved and yield the Lagrange-Farkas multipliers as well as restrictions for the entropy and the entropy flux. Applying these solutions of
the Liu equations, the dissipation inequality can be simplified to the following form
\begin{multline}
0\leq \sigma_s = \partial_i \mathfrak{\hat J}^i +
        \hat{q}^i\partial_i(\partial_e(\rho\mathfrak{s}))+
        \hat P^{ij}\partial_i\left(\partial_{v^j}\mathfrak{s}\right)+ \\
        \partial_jv^j\left(\hat s+e\partial_e\hat s -
                \rho \partial_\rho \hat{s}+
                \frac{\rho^{2}}{2}\partial_i\left(\partial_{
           \partial_i\rho} \mathfrak{s}\right) \right)+
\partial_jv^i\left(
       \frac{\rho^{2}}{2}\partial_i\left(\partial_{
           \partial_j\rho}\mathfrak{s}\right)\right).
\end{multline}
Here we have introduced the specific entropy as $\mathfrak{s}:=\hat s/\rho$. It is worth to give this inequality, the entropy production in a more traditional form, without indices, too
\begin{multline}
 \nabla\cdot{\bf J}_0
    + {\bf q}\cdot \nabla\partial_e(\rho \mathfrak{s})+\\ 
    \nabla \partial_{\bf v}  \mathfrak{s}:{\bf P}
    + \left[\frac{\rho^2}{2}\left(\nabla\cdot 
        \partial_{\nabla\rho}\mathfrak{s}{\bf I}
        + \nabla\partial_{\nabla\rho}\mathfrak{s}\right)
    +(\hat s +e\partial_e \hat{s}-\rho\partial_\rho\hat{s}) {\bf I}\right]:
    \nabla{\bf v} \geq 0.
\end{multline}

Here ${\bf I}$ denotes the second order unit tensor
$\delta_{ij}$, $\nabla$ is the gradient, $\nabla\cdot$ is the divergence of the corresponding field quantity. To get the traditional form of the equation we
should introduce the internal energy of the fluid as the difference of the total and kinetic energies and assume that the entropy function, does not depend on the total and kinetic energies independently, but only on the internal energy $\hat s=\hat{s}(\rho,\partial_i\rho,e-\rho v^2/2)$. Moreover, we want to define the entropy as an extensive quantity, therefore we require that the specific entropy depends on the specific internal energy $\epsilon$:
\begin{equation}
    \hat s(\rho,\partial_i\rho,{v}^i,e) =
    \rho \mathfrak{s}\left(\rho,\frac{e}{\rho}-\frac{v^2}{2},
    \partial_i\rho\right).
 \label{tradf_s}\end{equation} 

\noindent From this form of the entropy function we get the Gibbs relation 
$$
d\epsilon = Td\mathfrak{s}+\frac{p}{\rho^{2}}d\rho-A^id \partial_i\rho.
$$
where $\epsilon=\frac{e}{\rho}-\frac{v^2}{2}$ is the specific internal energy. The temperature and pressure are defined by the customary partial derivatives of the entropy. The temperature can be connected both to the derivative by the total energy and the internal energy, because the corresponding derivatives are equal. $A^i$ is defined as the partial derivative of the entróopy by the density gradient, like the traditional intensives. :$$
\partial_e(\rho\mathfrak{s})=\frac{1}{T},\quad
\partial_{v^j}\mathfrak{s}=-\frac{v^j}{T},\quad
\partial_\rho \mathfrak{s}= -\frac{p}{T\rho^2},\quad
\partial_{\partial_i\rho}\mathfrak{s}=\frac{A^i}{ T}.
$$

With these quantities we can write the dissipation inequality and the entropy flux as
\begin{multline}
0\leq \sigma_s = \partial_i \mathfrak{\hat J}^i +
        (\hat{q}^i-v_j\hat{P}^{ij})\partial_i\frac{1}{T}-\\ 
        \frac{1}{T}\left(\hat P^{ij}-
                \left(p+ \frac{T\rho^{2}}{2}\partial_k\left(\partial_{
           \partial_k\rho} \mathfrak{s}\right)
           \right)\delta^i_j
           - \frac{T\rho^{2}}{2}\partial_j\left(\partial_{
           \partial_i\rho}\mathfrak{s}\right)\right)\partial_iv^j.
\end{multline}
\begin{eqnarray}
\hat{J}^i &=& (\hat{q}^i-v_j\hat{P}^{ji})\frac{1}{T}+
  \frac{\rho}{2}\left(\partial_{\partial_i\rho}\hat s \partial_jv^j+
        \partial_{\partial_k\rho}\hat s \partial_kv^i\right) + 
  \mathfrak{\hat{J}}^i.
\end{eqnarray}

We can see that a flux of the internal energy is introduced as in case of the traditional, first order weakly nonlocal theories. However, the appearance of temperature inside the expression of the viscous pressure indicates the possibility of alternate, better definitions. 

Now we change the notation to a coordinate invariant one, introducing the nabla operator for the space derivatives  as it is customary in  fluid mechanics. In this case e.g. $\partial_i v^i= \nabla\cdot {\bf v}$ and $\partial_i v^j= \nabla{\bf v}$.   In the pure mechanical, reversible case our thermodynamic force for mechanical interactions is zero. Therefore we introduce the  {\em nonlocal reversible pressure} as
\begin{equation}
\hat{\bf P}^{r}=
    \frac{T\rho^2}{2}\left[\left(
        \nabla\cdot\partial_{\nabla\rho}\mathfrak{s}-
        2 \partial_\rho \mathfrak{s}\right){\bf I}
        + \nabla\partial_{\nabla\rho}\mathfrak{s}
    \right].
\label{P_rev}\end{equation}

If the pressure is equal to the reversible pressure, there is no dissipation, the theory is
reversible (conservative). In case of a local entropy (independent of
the gradient of the density) then we obtain
\begin{equation}
\hat{\bf P}\rev\Euler(\rho) := -T\rho^2 \drho \mathfrak{s}(\rho) {\bf I},
\end{equation}
\noindent therefore, the corresponding equations are of the ideal
Euler fluid, where $p(\rho) = -T\rho^2 \drho \mathfrak{s}(\rho)$ is the
scalar pressure function. 
Introducing the viscous pressure ${\bf P}\visc$ as usual, we can
solve the dissipation inequality and give the
corresponding Onsagerian  conductivity equation as
$$
\hat{\bf P}\visc := \hat{\bf P} - \hat{\bf P}\rev = {\bf L}_{ONS}\cdot \nabla{\bf
v}.
$$

Here ${\bf L}_{ONS}$ is a nonnegative constitutive function. Let
us recognize that if $\mathfrak{s}$ is independent of the gradient of
the density, ${\bf L}_{ONS}$ is constant, and $\hat{\bf P}\visc$ is an
isotropic function of only $\nabla{\bf v}$, then we obtain the
traditional Navier-Stokes fluid (see e.g. \cite{Gya70b}).

One can prove easily that the reversible part of the pressure  is {\em potentializable}, i.e., there is a
 scalar valued function $U$ such that
\begin{equation}
    \nabla\cdot \hat{\bf P}\rev = \rho\nabla \hat U.
\label{potcond}\end{equation}

\noindent $\hat U$ can be calculated from the entropy function as
\begin{equation}
    \hat U = \nabla\cdot(\rho\partial_{\nabla\rho}\mathfrak{s}) -
    \partial_\rho(\rho \mathfrak{s}).
\label{qpres_gen}\end{equation}

Therefore in case of reversible fluids the momentum balance can be
written alternatively as
\begin{equation}
\rho\dot{\bf v} + \nabla\cdot{\bf \hat P}^r = {\bf 0} \quad
\Longleftrightarrow \quad \dot{\bf v} + \nabla \hat U = {\bf 0}.
\label{balNew}\end{equation}

Let us give an interesting particular example of a weakly nonlocal fluid.

\subsection{Schr\"odinger-Madelung fluid} Here the entropy is
defined as
\begin{equation}
    s\SchM(\rho,\nabla\rho,{\bf v}) =
    -\frac{\nu}{2}\left(\frac{\nabla\rho}{2\rho}\right)^2
    - \frac{{\bf v}^2}{2} =
    -\frac{\nu}{8} (\nabla \ln\rho)^2 - \frac{{\bf v}^2}{2},
\label{Schf_s}\end{equation}

\noindent where $\nu$ is a constant scalar. The corresponding
reversible pressure is
\begin{equation}
{\bf P}\rev =
    - \frac{\nu}{8}\left(\Delta\rho {\bf I} + \nabla^2\rho
    - \frac{2 \nabla\rho\circ\nabla\rho}{\rho}\right),
\label{P_Sch}\end{equation}

\noindent where $\circ$ denotes the tensorial/dyadic product, as
mentioned before. The potential is
\begin{equation}
U\SchM = - \frac{\nu}{4\rho}\left(\Delta \rho -
        \frac{(\nabla\rho)^2}{2\rho}\right)
    = -\frac{\nu}{2} \frac{\Delta R}{R},
\label{U_B}\end{equation}

\noindent where we introduced $R=\sqrt{\rho}$ to show that (\ref{U_B}) is the quantum potential in the {de~Broglie}-Bohm
version of quantum mechanics (if $\nu = \hbar^2/m^2$)
\cite{Boh51b,Hol93b}.

The entropy flux of the Schr\"odinger-Madelung fluid is
\begin{equation}
    {\bf J}\SchM = -{\bf v}\cdot{\bf P}\rev
       - \frac{\nu}{8}(\nabla\rho\nabla\cdot{\bf v}
       + \nabla\rho\cdot\nabla{\bf v}).
\label{js_Sch}\end{equation}

\subsubsection{Remark} The results of this subsection are   strange with the traditional interpretation of objectivity and frame independence. That question will be discussed from a more general point of view in the next section. 

\subsubsection{Discussion} An important property of the Schr\"odinger-Madelung fluid is that
if the motion of the fluid is vorticity free, $\nabla\times{\bf v}
={\bf 0}$, then the mass and momentum balances can be transformed
into and united in the Schr\"odinger equation. Hence the balance
of momentum (\ref{cond_mom}) can be derived from a Bernoulli
equation (in a given inertial reference frame). Defining a scalar
valued phase (velocity potential) by
$$
{\bf v} = \frac{\hbar}{m} \nabla S,
$$
we obtain the Bernoulli equation observing that the second part
of (\ref{balNew}) is the gradient of 
\begin{equation}
\frac{\hbar}{m}\frac{\partial S}{\partial t} 
+\frac{{\bf v}^2}{2} - U\SchM =0.
\label{Bern_eq}\end{equation}

Then, introducing a single complex valued function $\psi := R e^{i
S}$ that unifies $R = \sqrt{\rho}$ and $S$, it is easy to find
that the sum of (\ref{cond_mass}) multiplied by $i \hbar e^{i S}/(2R)$
and (\ref{Bern_eq}) multiplied by $m R e^{i S}$ form together the
Schr\"odinger equation for free particles:
\begin{equation}
i \hbar \frac{\partial \psi}{\partial t} = -\frac{\hbar^2}{2 m}
\Delta\psi.
\label{Schro_eq}\end{equation}

It is remarkable that the structure of
quantum mechanics appears in a classical thermodynamic approach without any explicit distinctive assumption related to the microscopic quantum  world. In this sense the basic assumptions of our derivation are  rather weak and very general. 

\section{Summary and outlook}

In this paper we have shown that the Second Law of thermodynamics
provides a general, uniform, rigorous and constructive method to investigate weakly nonlocal extensions of classical and nonclassical non-equilibrium thermodynamics. We have seen how one can generate evolution equations for internal variables, to understand the constitutive structure of Classical Irreversible Thermodynamics and to restrict the pressure tensor of ideal and viscous Korteweg fluids.

In the suggested approach the choice of the constitutive state space is a physical question. The number of the necessary constraints for a given constitutive state space was introduced intuitively, according to our calculational experience, always looking forward to an explicit solution of the 
Liu equations. Cimmelli in \cite{Cim07a} introduces a different philosophy, 
requiring that the number of the constraints must equal the number of constitutive state variables. Additional gradients of the constraints are to be introduced in case of necessity. In general, the method given in \cite{Van05a} introduces less additional constraints, therefore the results are more easily applicable. On the other hand, the method proposed in \cite{Cim07a} is more cumbersome and the interpretation of the consequences of the Second Law can be less straightforward. However, it is based on a precise rule. 

One of the interesting observations was that we were able to derive some restrictions to the reversible part of the evolution equations. There we have encountered equations of Euler-Lagrange type in case of zero entropy production in the corresponding interaction (static Ginzburg-Landau, dynamic degrees of freedom, Bohm-potential for Korteweg pressure). In a sense we were able to derive   Hamiltonian variational principles, we have proved their existence from the Second Law of thermodynamics. 

There are several other classical and nonclassical weakly nonlocal equations of physics that could be investigated in this general frame. We have analyzed    weakly nonlocal extensions of the heat conduction equations (both Fourier and Cattaneo-Vernotte) were researched, too. In this respect we have got that special kind of internal variables, the so called {\em current multipliers} can  represent some kind of nonlocal effects in the theory. They can lead to theoretical structures like the Guyer-Krumhansl equation of heat conduction \cite{Van01a1}. This concept is originated in general extensions of the entropy flux \cite{Ver83a,Nyi91a1}.  These possibilities have been  investigated together with higher order weakly nonlocal extensions in case of generalized heat conduction  and generalized Ginzburg-Landau equations in \cite{CiaAta07a,Van05a1}. The flux hierarchy of extended irreversible thermodynamics and the origin of balance form evolution equations for internal variables was researched in \cite{CimVan05a,CimVan06p}. 

On the other hand, we have researched phase separated multicomponent fluids and got a structure with natural instabilities (kind of nonlocal phase boundaries) similar to the Goodman-Cowin pressure \cite{GooCow72a,Van04a,Van04a1}.

However, why should we restrict ourselves to weakly nonlocal extensions of the 
constitutive functions? Why do not we consider time derivatives in exploiting the Second Law? The answer to this question leads to one of the most fundamental problems of physics, the question of objectivity, in particular, to the question of material frame indifference. In this paper we restricted ourselves to evolution equations and physical quantities given in an inertial reference frame. However, we know well, that the laws of physics are independent of an external observer. In particular, the constitutive functions can depend only on the material (e.g. on the motion of the continua) but not on the motion of an external observer. There are strong arguments that the usual formulation of the material frame indifference and the usual concept of objectivity is wrong \cite{MatGru96a,MatVan06a,MatVan08p,Fre09a}. These investigations show well, that the core of the problem is in the usual intuitive concept of the nonrelativistic spacetime \cite{MatVan07a}. Therefore, we need more rigorous analysis of the kinematics of classical continua from this point of view and an absolute, frame independent formulation of the exploitation method of the Second Law. As regards the new foundations of finite deformation kinematics the investigations of F\"ul\"op seem to settle a good starting point \cite{Ful08a3e}.

The results of those investigations are exploited in two key points in this paper. First of all, from the objectivity point of view, considering  mathematical models of nonrelativistic spacetime, the successes of gradient extensions can be well understood. The gradient of a physical quantity is a spacelike component of a four-covector (spacetime quantity) and therefore it is independent of an observer \cite{Mat93b}: for inertial observers three-vectors are transforming by Galilei transformation, but three-covectors are invariant. Therefore \textit{gradients are objective}. On the other hand, for fluids we have started our investigations with an explicitly velocity dependent constitutive state space. That is excluded by the old concept of objectivity, because the relative three-velocities are frame dependent. However, it is not excluded according to the new concept of objectivity, because the four-velocities are independent of the frame. 
Our treatment of a one component fluid here is not a true objective treatment, from more than one point of view. However, we have done the first steps toward the development of objective exploitation methods of the Second Law, too. As the problem of objectivity is in a sense more easily understandable from the point of view of a relativistic spacetime model, we have investigated the nonequilibrium thermodynamics of special relativistic fluids \cite{Van08a}. Moreover, we have started to investigate some possible practical consequences of incorporating our generalization of the objective time derivatives of rheology into a thermodynamic framework \cite{VanAss06ae,Van08a1}.

Finally let us emphasize how surprising is our result regarding fluid mechanics. The original observation of Madelung was that the Schr\"odinger equation can be transformed into an interesting fluid mechanical form \cite{Mad26a}. The obervation of Bohm was that the same equation can be transformed into a Newtonian form \cite{Boh51b}. In both cases the Schr\"odinger equation is postulated, coming out of the blue. Here we have \textit{derived} the fluid mechanical equivalent of the Schr\"odinger equation in a very general framework, from a surprisingly minimal set of assumptions, from the basic balances, the Second Law of thermodynamics and nonlocality of the interactions were used. The existence of such derivation is a disturbing fact that is hardly understandable from the traditional point of view regarding the relation of the  irreversible macro- and reversible microworld.

\section*{Acknowledgements}

This research was supported by OTKA T048489. The author is grateful to T. Matolcsi and C. Papenfuss for the discussions and the corrections and for the valuable remarks of the referees.

\section{Appendix - Farkas's lemma and some of its consequences}

\begin{lem}{(Farkas)}
Let ${\bf a}_i\neq {\bf 0}$ be vectors in a  finite dimensional
vector space $\mathbb{V}$, $i=1...n$, and $S = \{ {\bf p} \in
\mathbb{V}^* | {\bf p}\cdot{\bf a}_i \geq 0, i=1...n \}$. The
following statements are equivalent for all ${\bf b} \in
\mathbb{V}$:

(i) $ {\bf p}\cdot {\bf b} \geq 0$, for all ${\bf p} \in S$.

(ii) There are nonnegative real numbers $\lambda_1, ...,
\lambda_n$ such that ${\bf b} = \sum^n_{i=1} \lambda_i {\bf a}_i$.
\end{lem}

Proof:

$(ii) \Rightarrow (i)$ ${\bf p}\cdot \sum^n_{i=1} \lambda_i {\bf
a}_i = \sum^n_{i=1} \lambda_i {\bf p} \cdot {\bf a}_i \geq 0$ if
${\bf p} \in S$.

$(i) \Rightarrow (ii)$ Let us consider a maximal, linearly independent subset 
${\bf a}_1, ..., {\bf a}_l$ of $S$.

Let $S_0 = \{ {\bf y} \in \mathbb{V}^* | {\bf y}\cdot{\bf a}_i =
0, i=1...l \}.$ Clearly $\emptyset \neq S_0 \subset S$.

If  ${\bf y}\in S_0$ then $-{\bf y}$ is also in $S_0$, therefore
${\bf y}\cdot{\bf b}\geq 0$ and $-{\bf y}\cdot{\bf b}\geq 0$
together. Therefore for all ${\bf y}\in S_0$ it is true that ${\bf
y}\cdot{\bf b}= 0$.
As a consequence ${\bf b}$ is in the linear subspace generated by $\{{\bf
a}_i\}$, that is  there are real numbers $\lambda_1,...,
\lambda_l$ such that ${\bf b} = \sum^l_{i=1} \lambda_i{\bf a}_i$.

Moreover, for all $k\in \{1,...,l\}$ there is a ${\bf p}_k\in
\mathbb{V^{*}}$ such that ${\bf p}_k\cdot{\bf a}_k = 1$ and ${\bf
p}_k\cdot{\bf a}_i = 0$ if $i\neq k$. Evidently, ${\bf p}_k\in S$ for all $k$,
therefore $0\leq {\bf p}_k\cdot{\bf b} = {\bf p}_k \cdot
\sum^l_{i=1}\lambda_i {\bf a}_i =\sum^l_{i=1} \lambda_i {\bf p}_k\cdot {\bf
a}_i = \lambda_k$ is valid for all $k$. Lastly, we can choose zero
multipliers for the vectors that are not independent.
$\blacksquare$

\begin{rem}
In the following the elements of $\mathbb{V}^*$ are called {\em
independent variables} and $\mathbb{V}^*$ itself is called the
{\em space of independent variables}. The inequality in the first
statement of the lemma is called {\em objective  inequality} and the
nonnegative numbers in the second statement are called {\em
Lagrange-Farkas multipliers}. The inequalities determining $S$ are
the {\em constraints}.

In the calculations an excellent reminder is to use Lagrange-
Farkas multipliers similarly to the Lagrange multipliers in case
of conditional extremum problems:
$$
{\bf p}\cdot{\bf b} - \sum^n_{i=1}\lambda_i {\bf p}\cdot
    {\bf a}_i
= {\bf p}\cdot({\bf b} - \sum^n_{i=1}\lambda_i \cdot {\bf
    a}_i) \geq 0,
\quad \forall {\bf p}\in \mathbb{V}^*
$$.

From this form we can read out the second statement of the lemma.
\end{rem}

\begin{rem}
The geometric interpretation of the theorem is important and
graphic: if the vector ${\bf b}$ does not belong to the cone
generated by the vectors ${\bf a}_i$, there exists a hyperplane
separating ${\bf b}$ from the cone.
\end{rem}

\subsection{Affine Farkas's lemma}

This generalization of the previous lemma was first published
simultaneously by A. Haar and J. Farkas as subsequent papers in
the same journal, with different proofs \cite{Haa18a,Far18a1}.
Later it was rediscovered  independently by others several times (e.g.
\cite{Neu??m,Sch98b}).

\begin{thm}{(Affine Farkas)} Let ${\bf a}_i\neq {\bf
0}$ be vectors in a finite dimensional vector space $\mathbb{V}$
and $\alpha_i$ real numbers, $i=1...n$ and  $S_A = \{ {\bf p} \in
\mathbb{V}^* | {\bf p}\cdot{\bf a}_i \geq \alpha_i, i=1...n \}$.
The following statements are equivalent for a ${\bf b}\in
\mathbb{V}$ and a real number $\beta$:

(i) ${\bf p}\cdot {\bf b} \geq \beta$, for all ${\bf p} \in S_A$.

(ii) There are nonnegative real numbers $\lambda_1,...,\lambda_n$
such that ${\bf b} = \sum^n_{i=1} \lambda_i {\bf a}_i$ and
$\beta\leq \sum^n_{i=1} \lambda_i \alpha_i$.
\end{thm}

Proof:

$(ii) \Rightarrow (i)$ ${\bf p}\cdot{\bf b} = {\bf p}\cdot
\sum^n_{i=1} \lambda_i {\bf a}_i = \sum^n_{i=1} \lambda_i {\bf p}
\cdot {\bf a}_i \geq \sum^n_{i=1} \lambda_i \alpha_i \geq \beta$.

$(i) \Rightarrow (ii)$ First we will show indirectly that the
first condition of Farkas's lemma is a consequence of the first
condition here, that is if (i) is true then ${\bf p}\cdot {\bf b}
\geq 0$, for all ${\bf p} \in S$.

Thus let us assume the contrary, hence there is ${\bf p}'\in S$,
for which ${\bf p}'\cdot{\bf b} < 0$. Take an arbitrary ${\bf
p}\in S_A$, then ${\bf p} + k{\bf p}' \in S_A$ for all real nonnegative 
numbers $k$. But now $({\bf p} + k{\bf p}')\cdot {\bf b} = {\bf
p}\cdot{\bf b} + k{\bf p}'\cdot{\bf b} < \beta$, if $k> \frac{{\bf
p}\cdot {\bf b}-\beta}{-{\bf p'}\cdot {\bf b}}$. That is a
contradiction.

Therefore, according to Farkas's lemma exist nonnegative Lagrange-Farkas multipliers
$\lambda_1,...,\lambda_n$ such that ${\bf b} = \sum^n_{i=1}
\lambda_i {\bf a}_i$. Hence $\beta \leq inf_{p\in S_A} \{{\bf
p}\cdot \sum^n_{i=1} \lambda_i{\bf a}_i \} = inf_{p\in S_A} \{
\sum^m_{i=1} \lambda_i{\bf p}\cdot {\bf a}_i \} = \sum^n_{i=1}
\lambda_i \alpha_i$. $\blacksquare$

\begin{rem} The multiplier form is a good reminder again
$$
({\bf p}\cdot{\bf b} - \beta) - \sum^m_{i=1}\lambda_i ({\bf
p}\cdot {\bf a}_i -\alpha_i) = {\bf p}\cdot({\bf b} -
\sum^m_{i=1}\lambda_i \cdot {\bf a}_i) - \beta  +
\sum^m_{i=1}\lambda_i \alpha_i \geq 0, \quad \forall {\bf p}\in
\mathbb{V}^*.
$$
\end{rem}

\begin{rem}
The geometric interpretation is similar to the previous one, but
with affine objects. If the line (one dimensional affine
hyperplane) $({\bf b},\beta)$ does not belong to the (affine) cone
generated by $({\bf a}_i,\alpha_i)$ there exists an affine
hyperplane separating ${\bf b}$ from the cone.
\end{rem}

\subsection{Liu's theorem}

Here the constraints are equalities instead of inequalities,
therefore the multipliers are not necessarily positive.

\begin{thm}{(Liu)}\label{Liuthm} Let ${\bf a}_i\neq {\bf 0}$ be vectors
in a finite dimensional vector space $\mathbb{V}$ and $\alpha_i$
real numbers, $i=1...n$ and $S_L = \{ {\bf p} \in \mathbb{V}^* |
{\bf p}\cdot{\bf a}_i = \alpha_i, i=1...n \}$. The following
statements are equivalent for a ${\bf b} \in \mathbb{V}$ and a
real number $\beta$:

(i) ${\bf p}\cdot {\bf b} \geq \beta$, for all ${\bf p} \in S_L$,

(ii) There are real numbers $\lambda_1,...,\lambda_n$ such that
\begin{equation}
    {\bf b} = \sum^n_{i=1} \lambda_i {\bf a}_i,
\label{Liu_eq}\end{equation}
and
\begin{equation}
    \beta\leq \sum^n_{i=1} \lambda_i \alpha_i.
\label{dis_ineq}\end{equation}
\end{thm}

Proof:

A straightforward consequence of the previous affine form of
Farkas's lemma because $S_L$ can be given in a form $S_A$ with the
vectors ${\bf a}_i$ and $-{\bf a}_i$, $i=1,...,n$: $S_L = \{ {\bf
p} \in \mathbb{V}^* | {\bf p}\cdot{\bf a}_i \geq \alpha_i, {\bf
p}\cdot(-{\bf a}_i) \geq \alpha_i, i=1...n \}$.

Therefore there are nonnegative real numbers
$\lambda_1^+,...,\lambda_n^+$ and $\lambda_1^-,...,\lambda_n^-$
such, that ${\bf b} = \sum^n_{i=1} (\lambda^+_i {\bf a}_i -
\lambda^-_i {\bf a}_i)  = \sum^n_{i=1} (\lambda^+_i - \lambda^-_i)
{\bf a}_i = \sum^n_{i=1} \lambda_i {\bf a}_i $ and $\beta \leq
\sum^n_{i=1} (\lambda^+_i \alpha_i - \lambda^-_i \alpha_i)$.
$\blacksquare$

\begin{rem} The multiplier form is a help in the applications
again
$$
0\leq  ({\bf p}\cdot{\bf b} - \beta) - \sum^n_{i=1} \lambda_i
    ({\bf p}\cdot {\bf a}_i -\alpha_i)
= {\bf p}\cdot({\bf b} - \sum^n_{i=1}\lambda_i \cdot {\bf a}_i)
    -\beta  + \sum^n_{i=1} \lambda_i \alpha_i,
\quad \forall {\bf p}\in \mathbb{V}^*.
$$
\end{rem}

\begin{rem} In the theorem with Lagrange multipliers for local
conditional extremum of differentiable function we apply exactly
the above theorem of linear algebra after a linearization of the
corresponding functions at the extremum point.
\end{rem}

Considering the requirements of the applications we generalize
Liu's theorem to take into account vectorial constraints:

\begin{thm}{(vector Liu)}
Let ${\bf A}\neq {\bf 0}$ in a tensor product $\mathbb{V}\otimes
\mathbb{U}$ of finite dimensional vector spaces $\mathbb{V}$ and
$\mathbb{U}$. Let \textbf{$\alpha$} $\in \mathbb{U}$ and $S_L = \{
{\bf p} \in \mathbb{V}^* | {\bf p}\cdot{\bf A} = {\bf \alpha}\}$.
The following statements are equivalent for a ${\bf b} \in
\mathbb{V}$ and a real number $\beta$:

(i) ${\bf p}\cdot {\bf b} \geq \beta$, for all ${\bf p} \in S_L$.

(ii) There is a vector ${\bf \lambda}$ in the dual of $\mathbb{U}$
such that
\begin{equation}
    {\bf b} = {\bf A}\cdot {\bf \lambda},
\label{Liu_veq}\end{equation}
and
\begin{equation}
    \beta\leq {\bf \lambda}\cdot {\bf \alpha}.
\label{dis_vineq}\end{equation}
\end{thm}

Proof:

Let us observe that we can get back the previous form of the
theorem by introducing a linear bijection ${\bf K}: \mathbb{U}
\rightarrow \mathbb{R}^n$, a {\em coordinatization} in
$\mathbb{U}$. Then by the vectors ${\bf a}_i := {\bf A}\cdot{\bf K}^* \cdot {\bf e}_i$, 
where ${\bf e}_i$ is the standard $i$-th unit vector in $\mathbb R^n$,
 and the numbers  $\alpha_i:={\bf K}\cdot \alpha$, we obtain that ${\bf b} =
\sum^n_{i=1} \lambda_i {\bf a}_i = {\bf A}\sum^n_{i=1}\lambda_i {\bf K}^*{\bf e}_i 
= {\bf A}\lambda $ where $\lambda:= \sum^n_{i=1}\lambda_i {\bf K}^*{\bf e}_i$.
$\blacksquare$

The previously excluded degenerate case of ${\bf A}={\bf 0}$
deserves a special attention. Then, of course, $\alpha=0$ and 
$S_L=\mathbb{V}^*$; this is, in fact, a degenerate case of all the previous theorems and evidently the following statements are equivalent 
for a ${\bf b} \in \mathbb{V}$ and a real number $\beta$:

(i) ${\bf p}\cdot {\bf b} \geq \beta$ for all ${\bf p}\in
\mathbb{V}^*$,

(ii) $\bf b=0$ and $\beta\le0$.

In this case the assignment of the vector space of \(\mathbb{V}^*\) is based on the inequality (i). We encountered various degeneracies in the calculations of the previous sections. 

\begin{rem}  In continuum physics and thermodynamics the above algebraic theorems are applied to differential equations and inequalities. There  the constraints are differential equations and therefore $\mathbb{V}$ is generated by derivatives of  some constitutive functions  in the differential equations. $\mathbb{V}^* $ is spanned by the {\em process directions}, the derivatives of the fields in the constitutive state space that  are not already there.  The
corresponding form of (\ref{Liu_veq}) and (\ref{dis_vineq}) are
called {\em Liu equation(s)} and the {\em dissipation inequality},
respectively.  \end{rem}

\end{document}